\documentstyle[12pt,epsf]{article}
\catcode`\@=11
\long\def\@makefntext#1{
\protect\noindent \hbox to 3.2pt {\hskip-.9pt  
$^{{\ninerm\@thefnmark}}$\hfil}#1\hfill}		

\def\@makefnmark{\hbox to 0pt{$^{\@thefnmark}$\hss}}  
	
\def\ps@myheadings{\let\@mkboth\@gobbletwo
\def\@oddhead{\hbox{}
\rightmark\hfil\ninerm\thepage}   
\def\@oddfoot{}\def\@evenhead{\ninerm\thepage\hfil
\leftmark\hbox{}}\def\@evenfoot{}
\def\sectionmark##1{}\def\subsectionmark##1{}}

\newcounter{sectionc}\newcounter{subsectionc}\newcounter{subsubsectionc}
\renewcommand{\section}[1] {\vspace*{0.6cm}\addtocounter{sectionc}{1} 
\setcounter{subsectionc}{0}\setcounter{subsubsectionc}{0}\noindent 
	{\normalsize\bf\thesectionc. #1}\par\vspace*{0.4cm}}
\renewcommand{\subsection}[1] {\vspace*{0.6cm}\addtocounter{subsectionc}{1} 
	\setcounter{subsubsectionc}{0}\noindent 
	{\normalsize\it\thesectionc.\thesubsectionc. #1}\par\vspace*{0.4cm}}
\renewcommand{\subsubsection}[1]
{\vspace*{0.6cm}\addtocounter{subsubsectionc}{1}
	\noindent {\normalsize\rm\thesectionc.\thesubsectionc.\thesubsubsectionc. 
	#1}\par\vspace*{0.4cm}}

\newcounter{appendixc}
\newcounter{subappendixc}[appendixc]
\newcounter{subsubappendixc}[subappendixc]

\renewcommand{\appendix}[1] {\vspace*{0.6cm}
        \refstepcounter{appendixc}
        \setcounter{figure}{0}
        \setcounter{table}{0}
        \setcounter{equation}{0}
        \renewcommand{\thefigure}{\Alph{appendixc}.\arabic{figure}}
        \renewcommand{\thetable}{\Alph{appendixc}.\arabic{table}}
        \renewcommand{\theappendixc}{\Alph{appendixc}}
        \renewcommand{\theequation}{\Alph{appendixc}.\arabic{equation}}
        \noindent{\bf Appendix \theappendixc #1}\par\vspace*{0.4cm}}

\renewenvironment{thebibliography}[1]
	{\begin{list}{\arabic{enumi}.}
	{\usecounter{enumi}\setlength{\parsep}{0pt}
\setlength{\leftmargin 1.25cm}{\rightmargin 0pt}
	 \setlength{\itemsep}{0pt} \settowidth
	{\labelwidth}{#1.}\sloppy}}{\end{list}}

\newcounter{itemlistc}
\newcounter{romanlistc}
\newcounter{alphlistc}
\newcounter{arabiclistc}

\newcommand{\fcaption}[1]{
        \refstepcounter{figure}
        \setbox\@tempboxa = \hbox{\footnotesize Fig.~\thefigure. #1}
        \ifdim \wd\@tempboxa > 6in
           {\begin{center}
        \parbox{6in}{\footnotesize\baselineskip=12pt Fig.~\thefigure. #1}
            \end{center}}
        \else
             {\begin{center}
             {\footnotesize Fig.~\thefigure. #1}
              \end{center}}
        \fi}

\newcommand{\tcaption}[1]{
        \refstepcounter{table}
        \setbox\@tempboxa = \hbox{\footnotesize Table~\thetable. #1}
        \ifdim \wd\@tempboxa > 6in
           {\begin{center}
        \parbox{6in}{\footnotesize\baselineskip=12pt Table~\thetable. #1}
            \end{center}}
        \else
             {\begin{center}
             {\footnotesize Table~\thetable. #1}
              \end{center}}
        \fi}

\def\@citex[#1]#2{\if@filesw\immediate\write\@auxout
	{\string\citation{#2}}\fi
\def\@citea{}\@cite{\@for\@citeb:=#2\do
	{\@citea\def\@citea{,}\@ifundefined
	{b@\@citeb}{{\bf ?}\@warning
	{Citation `\@citeb' on page \thepage \space undefined}}
	{\csname b@\@citeb\endcsname}}}{#1}}

\newif\if@cghi
\def\cite{\@cghitrue\@ifnextchar [{\@tempswatrue
	\@citex}{\@tempswafalse\@citex[]}}
\def\citelow{\@cghifalse\@ifnextchar [{\@tempswatrue
	\@citex}{\@tempswafalse\@citex[]}}
\def\@cite#1#2{{$\null^{#1}$\if@tempswa\typeout
	{IJCGA warning: optional citation argument 
	ignored: `#2'} \fi}}

 1
 1
 1

\font\ninerm=cmr9

\begin{document}
\pagestyle{empty}
\hfill{ANL-HEP-CP-96-34}
\vskip .5cm
\hfill{ May 2, 1996}
\vskip 1.cm
\centerline{\normalsize\bf THE PERTURBATIVE RESUMMED SERIES FOR 
TOP PRODUCTION\footnote{Presented by H. Contopanagos 
at the XXXI Rencontres de Moriond, ``QCD and High Energy Hadronic 
Interactions", March 23-30, 1996,
Les Arcs, France}} 
\vskip 1.cm
\begin{center}
Edmond L. Berger and Harry Contopanagos\\
High Energy Physics Division\\
Argonne National Laboratory\\
Argonne, IL 60439\\
\end{center}
\vskip .5cm
\centerline{\bf Abstract}
Our calculation of the total cross section for inclusive production of
$t\bar{t}$ pairs in hadron collisions is summarized.  The principal ingredient 
of this calculation is resummation of the  universal leading-logarithm effects
of gluon radiation to all orders in the quantum chromodynamics  coupling 
strength, restricted to the region of phase space that is  manifestly 
perturbative.  
We present predictions of the 
physical cross section as a function of top quark mass in proton-antiproton 
reactions at center-of-mass energies of 1.8 and 2.0 TeV.  

\section{Introduction}
\pagestyle{plain}
In this report we summarize our calculation of the inclusive cross
section for top quark production\cite{ref:edpapero}.
In hadron interactions at collider energies, the main production mechanisms 
for top-antitop quark ($t\bar{t}$) pair production, as modeled in perturbative 
quantum chromodynamics (pQCD), involve parton-parton collisions.  
  The gluonic radiative corrections
to the lowest-order channels create large enhancements of the partonic cross
sections near the top pair production threshold\cite{ref:dawson}.
The magnitude of the ${\cal O}(\alpha_s^3)$ corrections implies that fixed-order
perturbation theory will not necessarily provide reliable quantitative 
predictions for $t\bar{t}$ pair production at Fermilab Tevatron energies.  A 
resummation of the effects of gluon radiation to all orders in perturbation 
theory is called for in order to improve the reliability of the theory.  
\cite{ref:laeneno,ref:edpapero}.  

We use the 
notation $\alpha(\mu)\equiv \alpha_s(\mu)/\pi$, where $\mu$ is the common
renormalization/factorization scale of the problem. 
Unless otherwise specified, 
$\alpha\equiv \alpha(\mu=m)$ where $m$ is the mass of the top quark.
For the subprocess
\begin{equation}
i(k_1)+j(k_2)\rightarrow t(p_1)+{\bar t}(p_2)+g(k),
\label{subprocess}
\end{equation}
we use the partonic invariants\cite{ref:laeneno} 
\begin{equation}
s=(k_1+k_2)^2,\ t_1=(k_2-p_2)^2-m^2,\ u_1=(k_1-p_2)^2-m^2,\ (1-z)m^2=s+t_1+u_1.
\label{invariants}
\end{equation}
Through next-to-leading order (NLO), keeping only the 
leading logarithmic contributions, we write the total partonic cross section in 
 the $\overline{{\rm MS}}$ scheme as
\begin{equation}
\sigma_{ij}^{(0+1)}(\eta,m^2)=\int_{1-4(1+\eta)+4\sqrt{1+\eta}}^1dz
\left\{1+\alpha 2 C_{ij}\ln^2(1-z)\right\}\sigma'_{ij}(\eta,z,m^2)\ ,
\label{twelvep}
\end{equation}
where $\eta=s/(4m^2)-1$, $C_{ij}=C_F=4/3$ ($C_{ij}=C_A=3$) for $ij\equiv
q\bar{q}$ ($gg$),
$\sigma'_{ij}(\eta,z,m^2)\equiv {d\over dz}\bar{\sigma}_{ij}^B(\eta,z,m^2)$,
and $\bar{\sigma}_{ij}^B$ is the unpolarized Born partonic cross section.
We invoke universality with the Drell-Yan case ($l\bar{l}$ production).
Because the finite-order leading logarithms are identical in the $t\bar{t}$
and 
$l\bar{l}$ cases, we can resum them in $t\bar{t}$-production with the same 
function we use in the Drell-Yan case. 
We find 
\begin{equation}
\sigma_{ij}(\eta,m^2)=\int_{1-4(1+\eta)+4\sqrt{1+\eta}}^1
dz{\cal H}(z,\alpha)\sigma'_{ij}(\eta,z,m^2)\ .
\label{thirteen}
\end{equation}
The kernel of the hard part is\cite{ref:stermano} 
\begin{equation}
{\cal H}(z,\alpha)=1+\int_0^{\ln({1\over 1-z})}dx{\rm e}^{E(x,\alpha)}
\sum_{j=0}^\infty Q_j(x,\alpha)\ .
\label{fifteen}
\end{equation}
The kernel in Eq.~(\ref{fifteen}) depends solely on the resummation exponent
$E(x,\alpha)$, either explicitly, or through the functions $Q_j$
which depend exclusively on $E$. 

\section{The Resummation Exponent}

For the Drell-Yan
process, the exponent in moment space\cite{ref:stermanold} 
in the Principal Value Resummation 
(PVR) approach\cite{ref:stermano} may be written in 
the $\overline{{\rm MS}}$ factorization scheme as  

\begin{equation}
E(x,\alpha)=-g^{(1)}\int_P d\zeta{\zeta^{n-1}-1\over 1-\zeta}
\int_{(1-\zeta)^2}^1
{d\lambda\over \lambda}{\alpha\over 1+\alpha b_2\ln \lambda}\ .
\label{tonep}
\end{equation}
Here $P$ is a principal-value contour\cite{ref:stermano} and
$g^{(1)}=2C_{ij}$. 
It is important to note that Eq.~(\ref{tonep}) can include
in general all large logarithmic structures in the Drell-Yan case.
It has a perturbative
asymptotic representation\cite{ref:stermano}
\begin{equation}
E(x,\alpha)\simeq E(x,\alpha,N(t))=g^{(1)}\sum_{\rho=1}^{N(t)+1}\alpha^\rho
\sum_{j=0}^{\rho+1}s_{j,\rho}x^j\ ,
\label{teight}
\end{equation}
where 
\begin{equation}
s_{j,\rho}=-b_2^{\rho-1}(-1)^{\rho+j}2^\rho c_{\rho+1-j}(\rho-1)!/j! ,
\label{tnine}
\end{equation}
and  $\Gamma(1+z)=\sum_{k=0}^\infty c_k z^k$, where $\Gamma$ is the Euler gamma 
function.
This representation is valid in the moment-space interval 
\begin{equation}
1<x\equiv \ln n< t .
\label{tseven}
\end{equation}
This range of validity has the consequence that terms in the exponent of the
form $\alpha^k\ln^kn$ are of order unity, and terms with fewer powers
of logarithms, $\alpha^k\ln^{k-m}n$, are negligible.
This explains why resummation is completed in a finite number of steps. 
In addition, we discard monomials
$\alpha^k\ln^kn$ in the exponent because of the restricted
universality between the $t\bar{t}$ and $l\bar{l}$ processes.
The exponent we use in our calculations is
the truncation
\begin{equation}
E(x,\alpha,N)=g^{(1)}\sum_{\rho=1}^{N(t)+1}\alpha^\rho s_\rho x^{\rho+1} ,
\label{tseventeen}
\end{equation}
with the coefficients
$s_\rho\equiv s_{\rho+1,\rho}=b_2^{\rho-1}2^\rho/\rho(\rho+1)$.
The number of perturbative terms $N(t)$ in Eq.~(\ref{teight}) is
obtained by optimizing the asymptotic approximation 
$\bigg|E(x,\alpha)-E(x,\alpha,N(t))\bigg|={\rm minimum}$.
An excellent numerical approximation is provided
by the fit $ N(t)\simeq [t-3/2]$ ,
where the integer part is defined as the closest integer from either direction. 
Throughout this paper, we use the two-loop formula for the fixed coupling 
strength $\alpha(m)$.

In Fig. 1a we illustrate the validity of the asymptotic approximation 
for a value of t corresponding to $m=175$ GeV.  Optimization
works perfectly, with $N(t)=6$, 
and the plot demonstrates the typical breakdown of the asymptotic
approximation if $N$ were to increase beyond $N(t)$. This  
is the exponential rise of the infrared (IR) renormalons,
the $(\rho-1)!$ growth in the second term of Eq.~(\ref{tnine}). 
As long as $n$ is in the interval of Eq.~(\ref{tseven}),
all the members of the family in $n$ are optimized 
at the same $N(t)$, showing that the optimum number of 
perturbative terms is a function of $t$ only.

It is valuable to stress that we can derive the perturbative expressions,
Eqs.~(\ref{teight}), (\ref{tnine}) and (\ref{tseven}),
without the PVR prescription, although with less certitude.  
We begin with the {\it unregularized} form of Eq.~(\ref{tonep}), i.e., 
with the integral over $x$  on the real axis.
Expanding the inner integrand as a Taylor series around $\alpha$,
we find the same result as Eq.~(\ref{teight}), the only and 
major difference being
that we do not know the asymptotic properties of this series in 
the full range of moments $n$. 
For a fixed $t$ and $n$, one may use 
the monotonicity behavior of the corresponding partial
sums to try to determine an  
upper limit for the number of terms in Eq.~(\ref{teight}).  
This procedure is illustrated in Fig. 1b for $m=175$ GeV. We note that 
beyond a certain range of $N$, the\newpage 
\begin{figure}
\leftline{\hfill\hbox{\epsfxsize6.0cm\epsffile{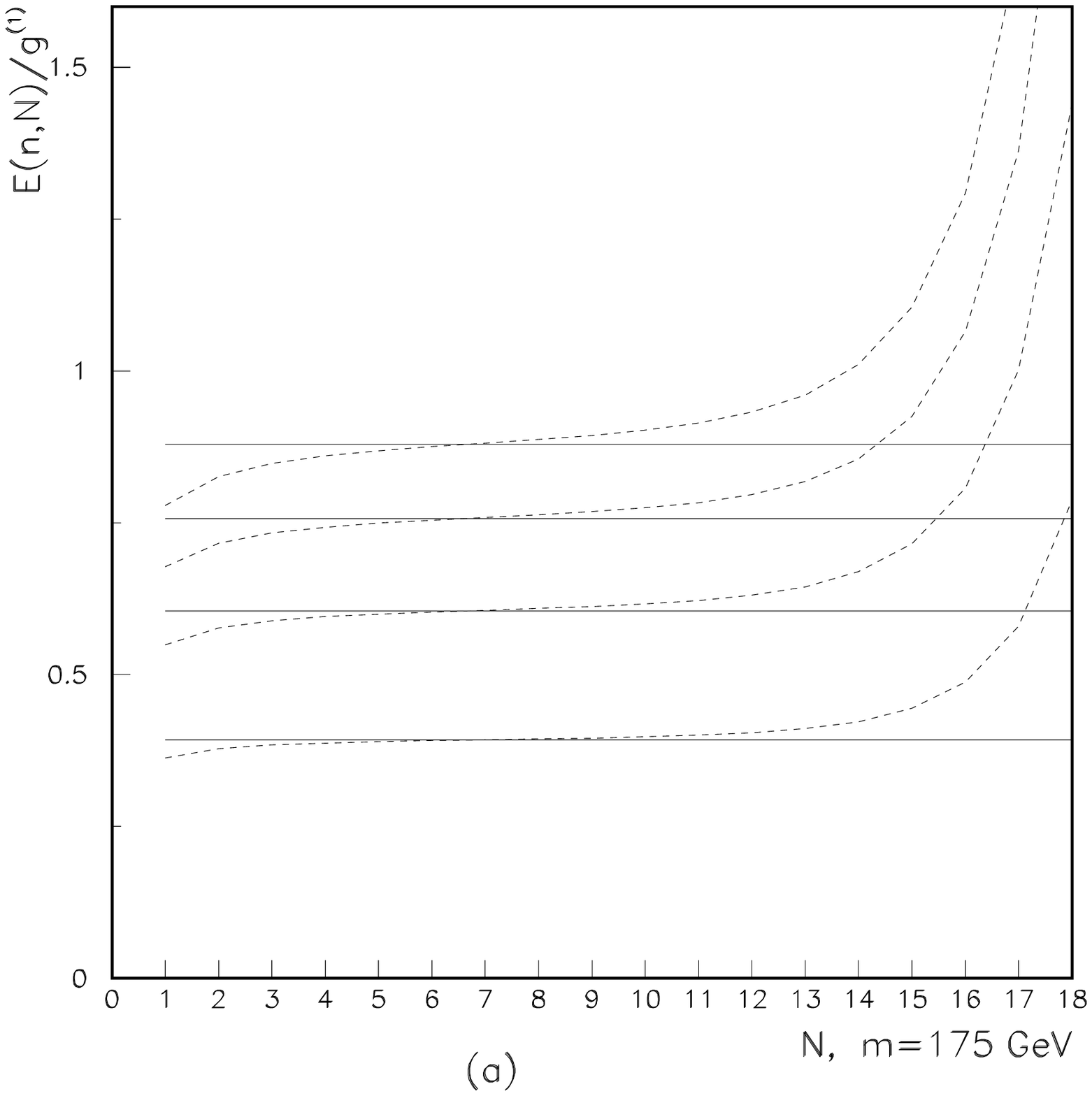}{\hskip 1.6cm}
\epsfxsize6.0cm\epsffile{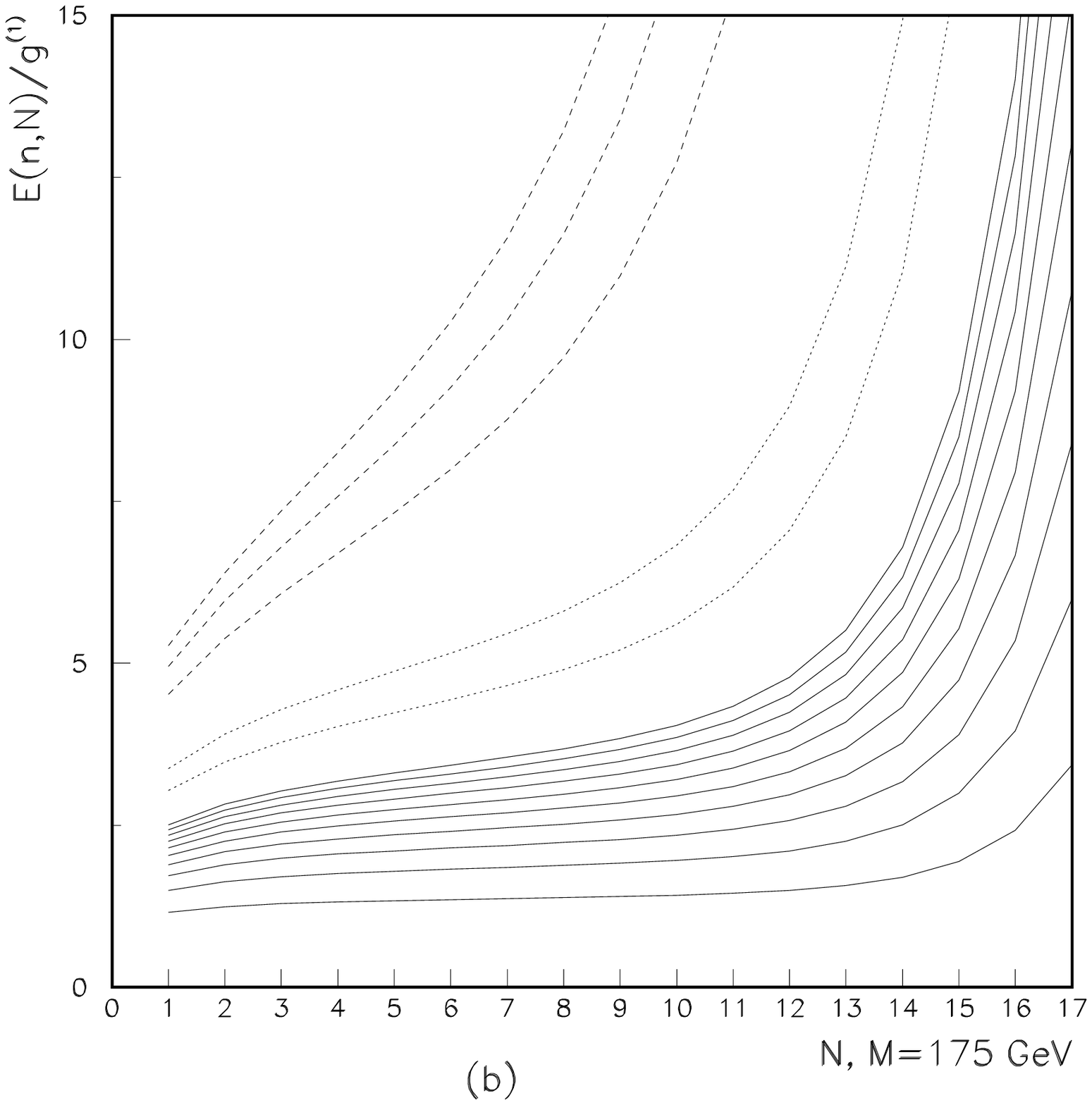}}\hfill}
\fcaption{Optimum number of perturbative terms in the exponent 
(a) with PVR  (solid family is for PVR, dashed for perturbative 
approximation, both families increasing, for parametric values 
$n=10,20,30,40$) and (b) using monotonicity of the partial sums (solid family
for $n\in \{10^2,10^3\}$ in steps of 100, dotted for $n=2,3\times 10^3$ and 
dashed for $n=1,2,3\times 10^4$). }
\end{figure}
exponent increases factorially, 
a demonstration of both the asymptotic nature of the series and
of the effect of the IR renormalons.  A range of 
optimum $N$ can be determined where the growth of the sum reaches a 
plateau, before the factorial growth sets in at large $N$.
The plateau is centered 
around $N_{opt}\simeq N(t)=[t-3/2]$, 
in agreement with PVR.

\section{The Resummed cross section}

The general form of the PVR cross section is given by Eqs.~(\ref{thirteen})
and  (\ref{fifteen}).
The functions $Q_j(x,\alpha)$ in Eq.~(\ref{fifteen}) appear
during the inversion of the Mellin transform\cite{ref:stermano}  
$n\leftrightarrow z$.
Specification  of  the perturbative regime in momentum
space follows from general expressions for the 
inversion of the Mellin transform and the meaning of the successive terms
in this inversion, once their perturbative approximations are used.
The functions $Q_0$ and $Q_1$,  can be calculated\cite{ref:stermano}: 
\begin{equation}
Q_0={1\over \pi}\sin(\pi P_1)\Gamma(1+P_1),\ \ 
Q_1\simeq 2\Gamma(1+P_1)P_2\cos(\pi P_1)\Psi(1+P_1)\ ,
\label{thseven}
\end{equation}
where $P_k=(\partial^k/k!\partial^kx)E$ and $
\Psi\equiv\Psi^{(0)}$ and $\Psi^{(k)}$ are the usual Polygamma
functions. 
For simplicity we include in the expression for  
$Q_1$ only terms that generate corrections
starting at ${\cal O}(\alpha)$.   
$Q_1$ contributes one less power of $x$ than $\alpha$ 
in the integrand of Eq.~(\ref{fifteen}), and it is formally subleading
relative to the contribution of $Q_0$. Nevertheless, from
Eq.~(\ref{thseven})  we see that this 
suppression is not true 
for values of $x$ such that $P_1(x,\alpha)\simeq 1$.  
We conclude that the 
perturbative region in momentum space is defined by
the inequality constraint
\begin{equation}
P_1(x_z,\alpha)\le 1\ ,\ \ x_z\equiv \ln\biggl({1\over 1-z}\biggr).
\label{theight}
\end{equation}
Since we intend to resum leading logarithms only, 
our main result for the perturbative resummed 
partonic cross section, denoted by $\sigma_{ij}^{R;pert}$, is
\begin{eqnarray}
& &\sigma_{ij}^{R;pert}(\eta,m^2)=\int_{1-4(1+\eta)+4\sqrt{1+\eta}}^{z_0}
dz\biggl[1+
\int_0^{x_z}dx{\rm e}^{E(x,\alpha)}
P_1(x,\alpha) 
\biggr]\sigma'_{ij}(\eta,z,m^2)\nonumber \\ 
& &=\int_{1-4(1+\eta)+4\sqrt{1+\eta}}^{z_0}
dz{\rm e}^{E(x_z,\alpha)}\sigma'_{ij}(\eta,z,m^2)\ ,
\label{thten}
\end{eqnarray}
where $z_0$ is the end-point calculated from Eq.~(\ref{theight}).
In order to achieve the best accuracy available we wish to include in 
our predictions as much as is known theoretically.  
Our ``final" resummed partonic cross section can therefore be 
written\cite{ref:edpapero}
\begin{equation}
\sigma^{pert}_{ij}(\eta, m^2,\mu^2)=\sigma^{R;pert}_{ij}(\eta,m^2,\mu^2)-
\sigma^{(0+1)}_{ij}(\eta,m^2,\mu^2)\Bigg|_{R;pert}+\sigma^{(0+1)}_{ij}(\eta,m^2,\mu^2)\ . 
\label{fthree}
\end{equation}
The second term is the part of the partonic cross section up to one-loop that is
included in the resummation, while the last term is the exact one-loop 
cross section\cite{ref:dawson}.

It is useful to translate our definition of the perturbative regime directly 
into a statement about the perturbative region in $\eta$.  
Our perturbative resummation probes the threshold down to the point
$\eta\ge \eta_0 =(1-z_0)/2 $.
Below this value, perturbation theory, resummed or otherwise, 
is not to be trusted. 
The physical cross section 
for each production channel is obtained through the factorization 
theorem.
\begin{equation}
\sigma_{ij}(S,m^2)={4m^2\over S}\int_0^{{S\over 4m^2}-1}d\eta\Phi_{ij}\biggl[
{4m^2\over S}(1+\eta),\mu^2\biggr]\sigma_{ij}(\eta,m^2,\mu^2) ,
\label{feleven}
\end{equation}
where the parton flux is 
$\Phi_{ij}[y,\mu^2]=\int_y^1{dx\over x}f_{i/h_1}(x,\mu^2)f_{j/h_2}(y/x,\mu^2)$.
The total physical cross section is obtained after incoherent addition of the 
contributions from  the the $q\bar{q}$ and $gg$ production channels.
We  ignore the small contribution from the $qg$ channel.  
We adopt one common scale $\mu$ for both the factorization and
renormalization scales.
A quantity of phenomenological interest is the differential cross section 
${d\sigma_{ij}(S,m^2,\eta)\over d\eta}$.
Its integral over $\eta$ is, of course, the total cross section.
In Fig. 2 we plot these distributions 
for $m=175$ GeV, ${\sqrt S}=1.8$ TeV and $\mu=m$.  
We observe that, at the energy of the Tevatron, resummation is 
significant for the $q\bar{q}$ channel and less so for the 
$gg$ channel. We show the total $t\bar{t}$-production\newpage 
\begin{figure}
\leftline{\hfill\hbox{\epsfxsize6.0cm\epsffile{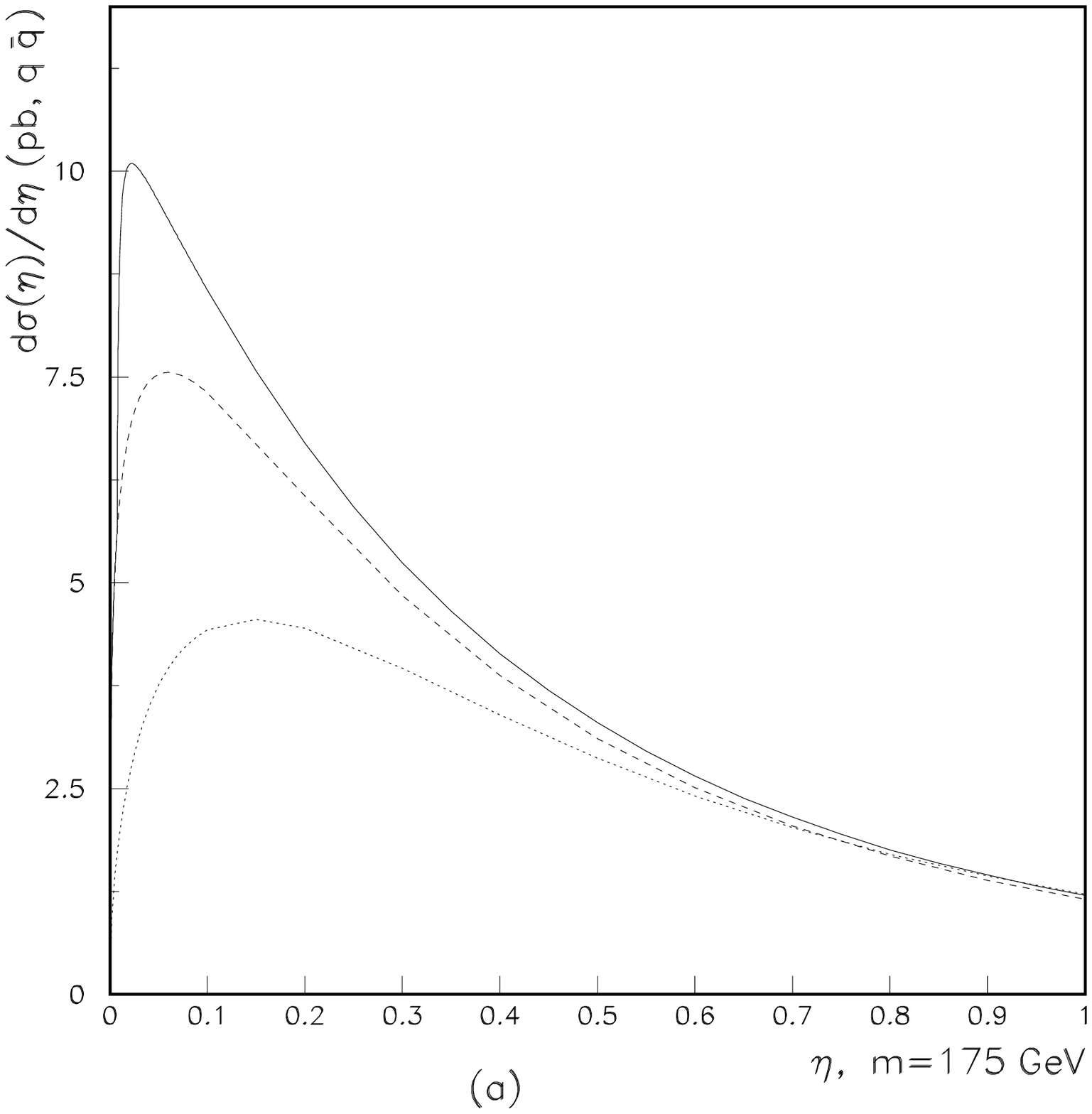}{\hskip 1.6cm}
\epsfxsize6.0cm\epsffile{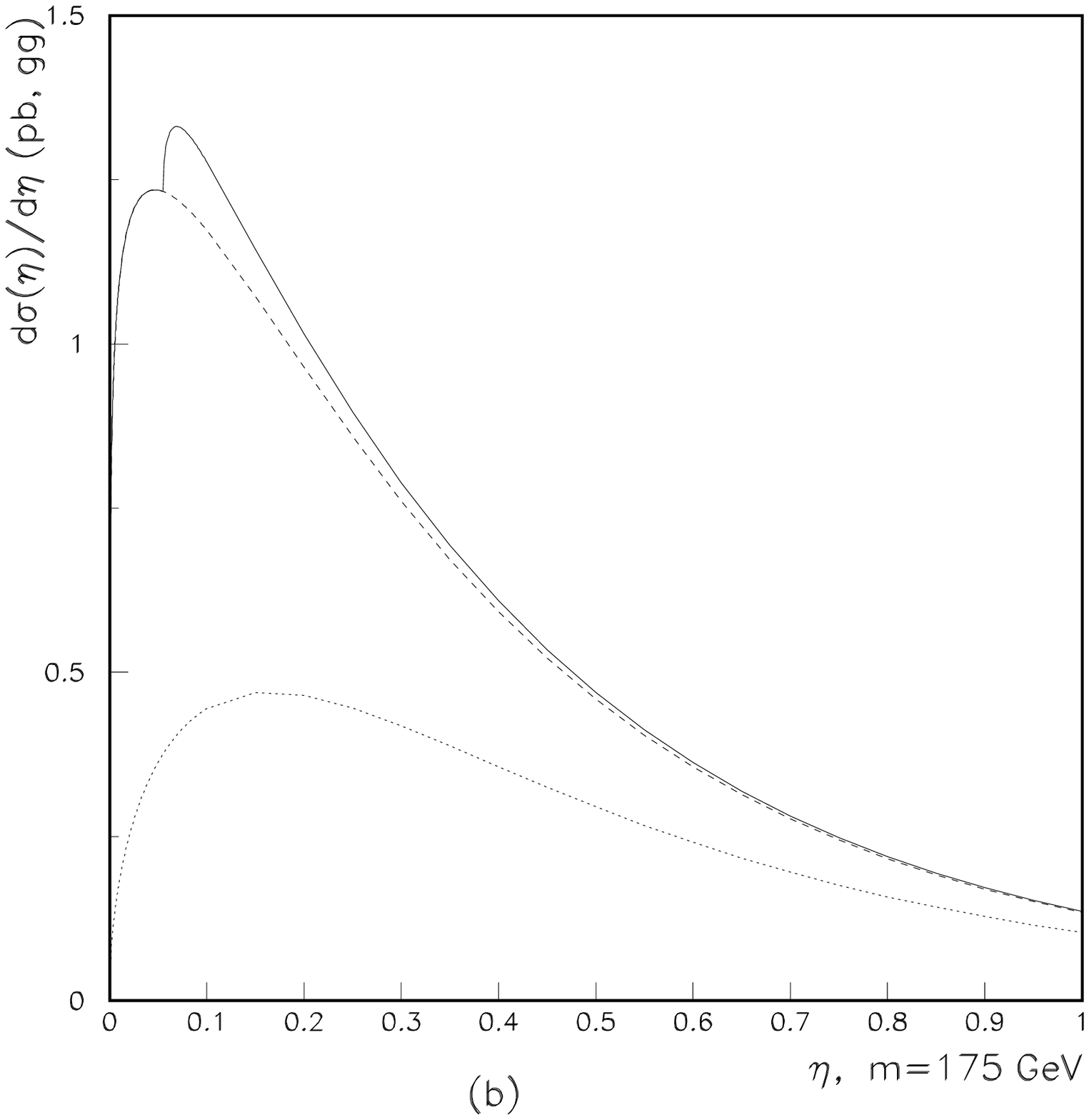}}\hfill}
\fcaption{Differential cross section $d\sigma/d\eta$ in the 
${\overline{\rm MS}}$-scheme for the (a) $q\bar{q}$  and (b) $gg$  channels:
Born (dotted),  NLO (dashed) and resummed (solid).}
\end{figure}
cross section 
as a function of top mass 
in Fig. 3. 
The central value of our predictions is obtained with the 
choice $\mu/m=1$, and the lower and upper limits are  the maximum 
and minimum of the cross section in the range of the hard scale 
$\mu/m\in\{0.5,2\}$. 
Our prediction of Fig. 3a is in  agreement with the 
 data\cite{ref:cdfdz}. We find
$\sigma^{t\bar{t}}(m=175\ {\rm GeV},\sqrt{S}=1.8\ {\rm TeV})=
5.52^{+0.07}_{-0.42}\ pb$.
In Fig. 3b we present our predictions for an upgraded Tevatron operating
at $\sqrt{S}=2$ TeV. Our cross section is larger than the NLO
one by about $9\%$.  We predict
$\sigma^{t\bar{t}}(m=175\ {\rm GeV},\sqrt{S}=2\ {\rm TeV})=
7.56^{+0.10}_{-0.55}\ pb$.
Over the range $\mu/m\in\{0.5,2\}$, the band
of variation of the strong coupling strength $\alpha_s$ is a
generous $\pm10$\%\ at $m$ = 175 GeV. 
\begin{center}
\begin{tabular}{|  c | c | c | c  |}\hline
$\sigma_{q\bar{q}}$($m=150$ GeV; ${\rm DIS}$) & $\mu/m=0.5$ & $\mu/m=1$ & $\mu/m=2$ \\ \hline\hline
NLO & 9.42 & 9.31 & 8.57 \\ \hline
$\sigma_{q\bar{q}}^{pert}$ & 9.76 & 9.92 & 9.31 \\ \hline
LSvN($\mu_0=0.1$m) & 7.9 & 10.0 & 9.7 \\ \hline\hline
$\sigma_{gg}$($m=150$ GeV; $\overline{{\rm MS}}$) & 
$\mu/m=0.5$ & $\mu/m=1$ & $\mu/m=2$ \\ \hline\hline
NLO & 2.51 & 2.22 & 1.81 \\ \hline
$\sigma_{gg}^{pert}$ & 2.53 & 2.30 & 1.89 \\ \hline
LSvN($\mu_0=0.2$m) & 1.76 & 4.38 & - \\ \hline
\end{tabular}
\end{center}
\tcaption{Physical cross sections in pb  at
$m=150$ GeV and $\sqrt{S} = 1.8$ TeV.  The corresponding LSvN predictions are 
also shown, where $\mu_0$ is their IR cutoff.}
Comparing with Laenen, Smith, 
and vanNeerven (LSvN)\cite{ref:laeneno},  
we find our central values are $10-14\%$ larger, 
and our estimated theoretical uncertainty is 
$9-10\%$ compared with their\newpage 
\begin{figure}
\leftline{\hfill\hbox{\epsfxsize6.0cm\epsffile{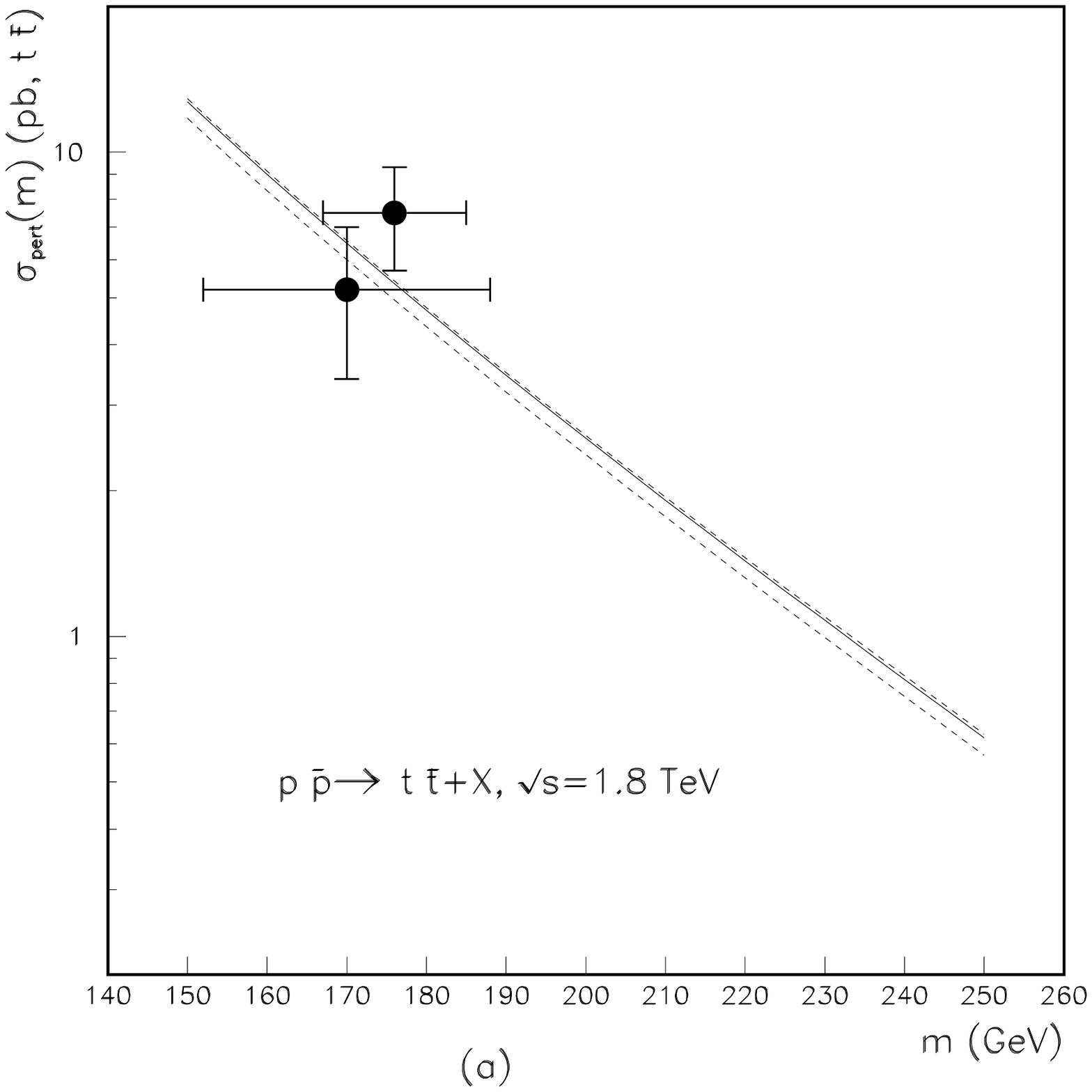}{\hskip 1.6cm}
\epsfxsize6.0cm\epsffile{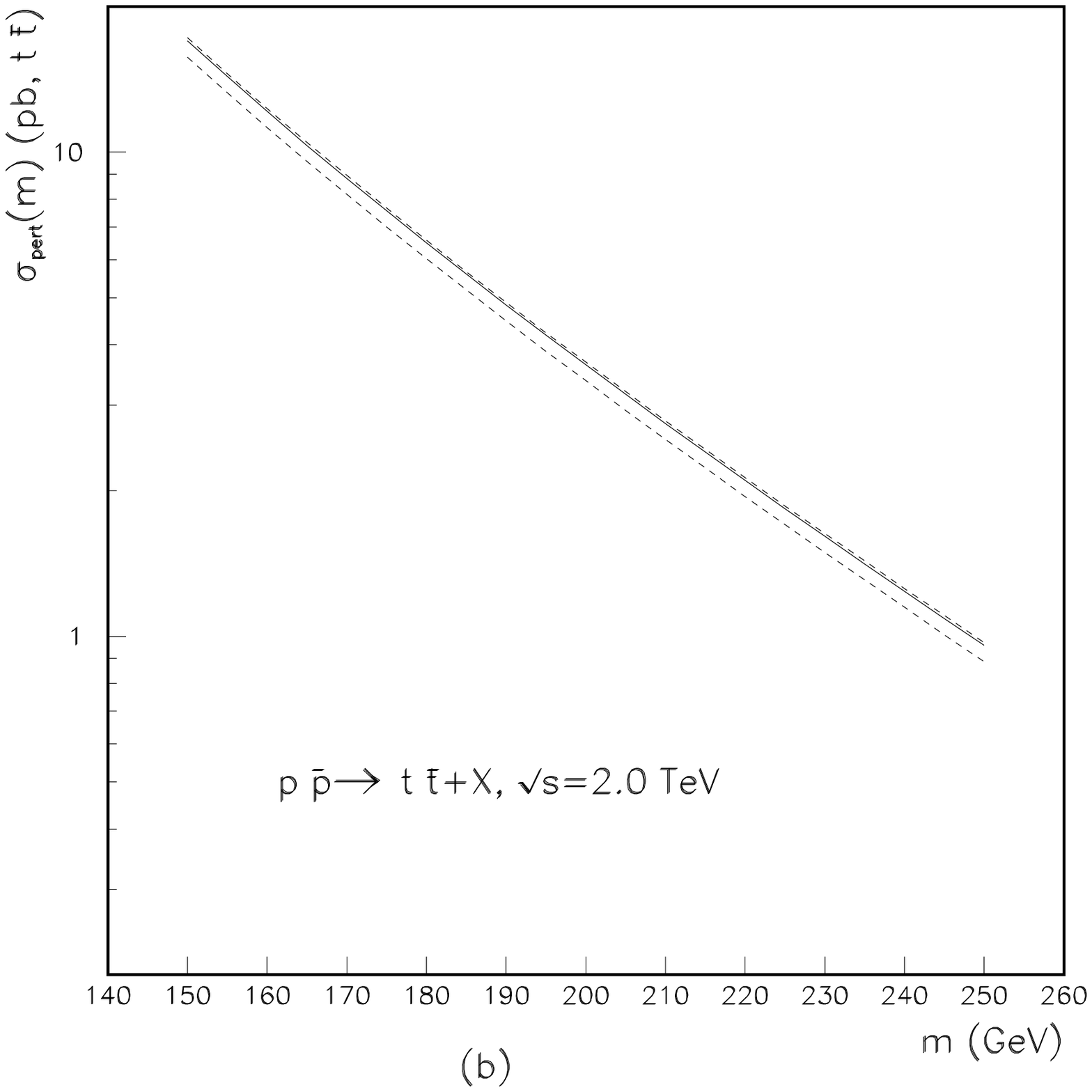}}\hfill}
\fcaption{Inclusive cross section for top quark production in the 
${\overline{\rm MS}}$-scheme. The dashed curves show our perturbative
uncertainty band, while the solid curve is our central prediction:
(a) $\sqrt{S}=1.8$ TeV  and (b)  $\sqrt{S}=2$ TeV.}
\end{figure}
$28\%-20\%$.  
Our Born cross section, however,  is about $3-5\%$ larger than the 
LSvN Born cross section due to the different parton distributions used in the
two calculations, including differences in $\Lambda$ which alone account 
for half or more of this increase.  
The two 
predictions have overlapping uncertainties and are, in this sense, in 
agreement. It is important to stress that the theoretical uncertainties are
estimated in quite different ways in the two methods.   
We use the standard $\mu$-variation, whereas LSvN obtain their uncertainty
primarily from variations of their undetermined IR cutoffs.
 One of the advantages 
of a resummation calculation should be diminished dependence
of the cross section on $\mu$, less variation than is present in fixed-order
calculations.  Table 1 shows that our resummed cross sections satisfy the test of 
stability under variation of the hard scale $\mu$.  The resummed results show
less variation than the NLO cross section. On the other 
hand, this is not true of the resummation of LSvN.
As shown in Table 1
their $q\bar{q}$ cross section in the DIS scheme has a $\mu$-variation of
$21\%$. For comparison, the NLO cross section shows a 
variation of $9\%$ and our resummed cross section a variation of $6\%$. 

Scheme dependence is an extra source of theoretical uncertainty, but it 
should produce minimal differences for physical
cross sections if these are calculated in an unambiguous resummation 
approach.  
To check for possible scheme 
dependent uncertainty, we perform our resummation for the dominant 
$q\bar{q}$ channel in both schemes.  The cross sections presented in Table 2 
show that our scheme dependence is insignificant, resulting in 
a difference of about $4\%$ for the cross section.\newpage  
\begin{center}
\begin{tabular}{|  c | c | c | c | c  |}\hline
$m$ (GeV)& $\sigma_{q\bar{q}}$({\rm DIS}; $\overline{{\rm MS}}$) & $\mu/m=0.5$ & $\mu/m=1$ & $\mu/m=2$ \\ \hline\hline
150 & NLO & 9.42; 9.68 & 9.31; 9.53 & 8.57; 8.73 \\ \hline
    & $\sigma_{q\bar{q}}^{pert}$ & 9.76; 10.16 & 9.92; 10.42 & 9.31; 9.87 \\ \hline
175 & NLO & 4.46; 4.54 & 4.39; 4.43 & 4.01; 4.02 \\ \hline
    & $\sigma_{q\bar{q}}^{pert}$ & 4.63; 4.78 & 4.69; 4.87 & 4.37; 4.58 \\ \hline
200 & NLO & 2.20; 2.21 & 2.15; 2.14 & 1.96; 1.93 \\ \hline
    & $\sigma_{q\bar{q}}^{pert}$ & 2.29; 2.34 & 2.30; 2.37 & 2.14; 2.21 \\ \hline
\end{tabular}
\end{center}
\tcaption{Physical cross sections in pb for the $q\bar{q}$ channel: DIS versus 
$\overline{{\rm MS}}$ scheme.}
 
\section{Discussion and Conclusions}

Our theoretical analysis shows that perturbative resummation without
a model for non-perturbative behavior is both  possible and 
advantageous.  In perturbative resummation, the perturbative region of phase
space is separated cleanly from the region of non-perturbative behavior.
The former is the region where large threshold corrections exponentiate
but behave in a way that is {\it perturbatively stable}.
The asymptotic character of the QCD perturbative series, including 
large multiplicative color factors, is flat, and excursions 
around the optimum number of perturbative terms does not create
numerical instabilities or intolerable scale-dependence.  Infrared
renormalons are far away from the stability plateau and, even though
their presence is essential for defining this plateau, they are 
of no numerical consequence in the perturbative regime. 
Large color factors, which are multiplicative, enhance the IR renormalon
effects and contribute significantly to limiting the perturbative regime.

Our resummed cross sections are about $9\%$ above the NLO
cross sections computed with the same parton distributions. The 
scale dependence of our cross section 
is fairly flat, resulting in a $9-10\%$ theoretical uncertainty.  This 
variation is smaller than the corresponding dependence of the NLO
cross section, and it is much smaller than the corresponding dependence 
of the resummed cross section of LSvN. 
There are other perturbative uncertainties, such as dependence on 
parton distributions and factorization scheme, which
affect our cross section minimally, at level of $4\%$ or less.
These variations are strongly correlated, so 
we opt not to add them in estimating the theoretical uncertainty.
Commenting briefly on  recent papers\cite{ref:catani}, 
in which the authors
state  that the increase in cross section they find with 
their resummation method
is of the order of $1\%$ over NLO, we 
stand firmly by our results. As we explained\cite{ref:edpapero} 
there is no actual 
factorial growth in our expressions, of the type they suggest, {\it precisely} because
we stay away from infrared renormalons in the exponent by optimizing
the number of perturbative terms, {\it and}
because our phase space is constrained within the perturbative regime 
 derived by controlling the 
non-universal subleading logarithms.
The numerical difference in the two approaches boils down 
to the treatment of the subleading logarithms, which can 
easily shift the results by a few percent, if proper care is not
taken. Our approach includes the universal leading
logarithms only while theirs includes non-universal subleading
structures which produce the suppression they find. In our opinion,  
their treatment of the subleading structures is not correct. 
We will present an  account of these
issues in a future publication.

Our theoretical analysis and the stability of our cross sections under $\mu$
variation provide confidence that our perturbative resummation procedure 
yields an
accurate calculation of the inclusive top quark cross section at Tevatron 
energies and exhausts present understanding of the perturbative content 
of the theory. Our prediction agrees with data, within the large 
experimental uncertainties. 

This work was supported by the US Department of Energy, Division 
of High Energy Physics, Contract W-31-109-ENG-38.

\end{document}